\documentclass[12pt,preprint]{aastex}
\sfcode`A=1000 \sfcode`B=1000 \sfcode`C=1000 \sfcode`D=1000
\sfcode`E=1000 \sfcode`F=1000 \sfcode`G=1000 \sfcode`H=1000
\sfcode`I=1000 \sfcode`J=1000 \sfcode`K=1000 \sfcode`L=1000
\sfcode`M=1000 \sfcode`N=1000 \sfcode`O=1000 \sfcode`P=1000
\sfcode`Q=1000 \sfcode`R=1000 \sfcode`S=1000 \sfcode`T=1000
\sfcode`U=1000 \sfcode`V=1000 \sfcode`W=1000 \sfcode`X=1000
\sfcode`Y=1000 \sfcode`Z=1000

\newcommand{\js}{}
\newcommand{\etal}{{\js et~al.\/}}
\newcommand{\eg}{{\js e.g.\/}}
\newcommand{\ie}{{\js i.e.\/}}
\newcommand{\Lsol}{$L_{\odot}$}
\newcommand{\Msun}{\mbox{$M_\sun$}}
\newcommand{\Lsun}{\mbox{$L_\sun$}}
\newcommand{\0}{\phantom{0}}  % for spacing in tables and such
\newcommand{\no}{\nodata}

\newcommand{\sst}{{\it Spitzer Space Telescope}}
\newcommand{\s}{{\it Spitzer}}

\slugcomment{v2 \today}
\shorttitle{The 8.0~\micron\ Luminosity Function}
\shortauthors{Huang et al.}
\received{2007 January 24}
\begin{document}
\title{The Local  Galaxy 8\,\micron\ Luminosity Function}
\author{J.-S.~Huang,$\!$\altaffilmark{1}
M. L. N. Ashby,$\!$\altaffilmark{1}
P.~Barmby,\altaffilmark{1}
M.~Brodwin,\altaffilmark{2}
M.~J.~I.~Brown,\altaffilmark{3}
N.~Caldwell,\altaffilmark{1}
R.~J.~Cool,\altaffilmark{4}
P.~Eisenhardt,\altaffilmark{2}
D.~Eisenstein,\altaffilmark{4}
G.~G.~Fazio,\altaffilmark{1}
E.~Le Floc'h,\altaffilmark{4}
P.~Green,\altaffilmark{1}
C.~S.~Kochanek,\altaffilmark{6}
N.~Y.~Lu,\altaffilmark{5}
M.~A.~Pahre,\altaffilmark{1}
D.~Rigopoulou,\altaffilmark{7}
J.~L.~Rosenberg,\altaffilmark{1,8}
H.~A.~Smith,\altaffilmark{1}
Z.~Wang,\altaffilmark{1}
C.~N.~A. Willmer,\altaffilmark{4}
\& 
S.~P.~Willner\altaffilmark{1}
}

\altaffiltext{1}{Harvard-Smithsonian Center for Astrophysics, 60 Garden Street,
Cambridge, MA 02138}
\altaffiltext{2}{Jet Propulsion, California Institute of Technology,
  Mail Stop 169-506, Pasadena, CA91109} 
\altaffiltext{3}{School of Physics, Monash  
University, Clayton 3800, Victoria, Australia} 
\altaffiltext{4}{Steward Observatory, University of Arizona, 933 N
  Cherry Ave., Tucson, AZ 85121} 
\altaffiltext{5}{Spitzer Science Center,
Caltech, 1200 E. California, Pasadena, CA 91125}
\altaffiltext{6}{Department of Astronomy, The Ohio State University,
  140 West 18th Ave., Columbus, OH 43210} 
\altaffiltext{7}{Department of Astrophysics, Oxford University, Keble
  Rd, Oxford, OX1 3RH, U.K.} 
\altaffiltext{8}{Current address: Department of Physics and 
  Astronomy, George Mason University, 4400 University Drive, MS 3F3,
  Fairfax, VA 22030}
\begin{abstract}
A \sst\ survey in the NOAO Deep-Wide Field in Bo\"otes  provides a
complete, 8~\micron-selected sample of galaxies to a limiting (Vega)
magnitude of 13.5.  In the 6.88~deg$^2$ field sampled, 79\% of the
4867 galaxies have spectroscopic redshifts, allowing an accurate
determination of the local ($z<0.3$) galaxy luminosity
function. Stellar and dust emission can be separated on the basis of
observed galaxy colors.  Dust emission (mostly PAH)
accounts for 80\% of the 8~\micron\ luminosity, stellar
photospheres account for 19\%, and AGN emission accounts for roughly
1\%. A sub-sample of the 8~\micron-selected galaxies have blue,
early-type colors, but even most of these have significant PAH
emission. The luminosity functions for the total 8~\micron\
luminosity and for the dust emission alone are both well fit by
Schechter functions. For the 8~\micron\ luminosity function, the
characteristic luminosity is $\nu L_{\nu}^*(8.0~\micron) = 1.8 \times
10^{10}$~\Lsun\ while for the dust emission alone it is $1.6 \times
10^{10}$~\Lsun\null.  The average 8~\micron\ luminosity density at
$z<0.3$ is $3.1 \times 10^7$~\Lsun~Mpc$^{-3}$, and the average luminosity
density from dust alone is $2.5 \times 10^7$~\Lsun~Mpc$^{-3}$.  This luminosity
arises predominantly from galaxies with 8~\micron\ luminosities ($\nu
L_{\nu}$) between $2\times 10^9$ and $2\times10^{10}$~\Lsun, i.e.,
normal galaxies, not LIRGs or ULIRGs.

\end{abstract}
\keywords{cosmology: observations --- galaxies:dust emission --- galaxies: 
Survey --- galaxies: mid-infrared}


\section{INTRODUCTION}

Galaxy emission in the mid-infrared (3\micron--10\micron) arises from
three main mechanisms \citep[\eg][]{Piovan2006}: 1) the
Rayleigh-Jeans tail of stellar emission; 2) emission from heated dust
in the interstellar medium \citep[ISM,][]{ces96, roe96, ver96,
ver01}; and 3) a power law component powered by accreting black holes
\citep{moo86, roc91, Elvis1994}. The ISM component displays
prominent, broad emission features at 3.3, 6.2, 7.7, 8.6, and
11.3~\micron\ \citep{gil73,Willner1977,Phillips1984} often attributed
to polycyclic aromatic hydrocarbons (PAH, \citealt{leg84, all85,
pug89}). Elliptical galaxies typically have little or no dust, and
therefore their mid-infrared SEDs are expected to be dominated by
stellar emission \citep{lu00, lu03, pah04a, pah04b}.  Star-forming
galaxies, including the extreme Luminous Infrared Galaxies (LIRG,
$L_{IR} > 10^{11}$~\Lsun) and Ultra-Luminous Infrared Galaxies
(ULIRG, $L_{IR} > 10^{12}$~\Lsun), show the PAH emission
features \citep{lu03,pah04a,eng05,gen98,rig99,Armus2007}.  Active nuclei can
occur in any galaxy type.  Galaxy spectral energy distributions
(SEDs) are the sum of radiation from all three mechanisms, and
because the proportions vary widely, galaxy SEDs in the mid-infrared
are very diverse.

Mid-infrared dust emission luminosity is a good indicator of a
galaxy's star formation rate.  It is insensitive to dust obscuration,
and it can be measured for large samples of galaxies at both
$z\approx0$ and $z\approx2$.  \citet[and earlier papers]{Roche1985}
showed that mid-infrared feature emission is prominent in star
forming galaxies, and ISO spectroscopic studies showed
that PAH feature strength increases with other measures of star
formation rate  \citep[SFR --- \eg][]{Vigroux1999,elb02}.
\citet{FS2004} argued for a quantitative 
relationship between PAH strength and SFR, and \citet{Wu2005}
confirmed a relationship using \s\ IRAC 8~\micron\ data to measure
the PAH emission.

Although PAH strength is a good SFR indicator, it is not perfect.
Galaxies with powerful AGN may have very weak or no PAH emission
features visible in their SEDs \citep{moo86, roc91, gen98,lau00,
ega04, alo05}. This doesn't mean active star formation is absent
because, for example, intense ultraviolet radiation produced by
the accreting black hole may dissociate PAH \citep{rig99}, especially
near the nucleus.
PAH features are also weak or absent in the SEDs of galaxies with low
metallicities \citep{eng05,Rosenberg2006,Wu2006,Madden2006} or low
luminosities \citep{hog05}.  At an extreme, \citet{hou04} found that
the SED of the blue compact galaxy SBS~0335$-$052 shows no PAH emission
at all. One possible explanation is that the formation and existence
of PAH depends on metallicity.  
Whatever the limitations of PAH emission as an SFR indicator, it is
still important to know how much PAH emission there is in the local
Universe and what kinds of galaxies that emission comes from. The
results should at least give the SFR from typical galaxies, if not
from unusual galaxy types.  Understanding the local population is an
essential step in determining how the SFR has evolved with time.

The \sst\ offers the opportunity to observe large samples of galaxies
at 8~\micron.  At low redshift, {\it Spitzer's\/} Infrared Array
Camera (IRAC, Fazio et al. 2004) can detect the 6.2, 7.7, and
8.6~\micron\ PAH emission features with its 5.8 and 8.0~\micron\
detector arrays, and IRAC's wide field-of-view ($5' \times 5'$) and
high sensitivity allow IRAC to map large areas of sky efficiently.
Maps at 24~\micron\ made with the Multi-band Imaging Photometer
\citep[MIPS ---][]{Rieke2004} are an ideal way of selecting large
samples of galaxies with PAH emission at $z \sim 2$ \citep{pap04,
yan04, hou05, pap06, web06, Caputi2006}, and many authors
\citep[eg][]{hou05, yan05, luz05, Lagache2004} have used the Infrared
Spectrograph \citep[IRS ---][]{hou04} to study the PAH emission
features in the high-redshift $2 \la z \la 3$ population.

This paper focuses on the 7.7 plus 8.6~\micron\ PAH emission features
from galaxies at low redshift detected in the IRAC 8.0~\micron\
band. By combining redshifts and IRAC photometry, we obtain an IRAC
8.0~\micron-selected local galaxy sample and derive a local PAH
luminosity function. The IRAC 8.0~\micron-selected galaxy sample is
selected at the same rest wavelength as a MIPS 24~\micron-selected
sample at $z=2$ \citep[e.g.,][]{Caputi2006} or an {\it AKARI}
16~\micron-selected sample at $z=1$.  A 
comparison between these populations can thus reveal the effects of
evolution  between $z=2$ and the present.  The paper is
arranged as follows: \S 2 describes the IRAC data and the sample
selection.  \S 3 discusses K-corrections for the IRAC bands and shows
the color-luminosity relations of the sample galaxies. The luminosity
function is presented in \S 4, and \S 5 summarizes the results.
Distances throughout this paper are based on $H_0 =
70$~km~s$^{-1}$~Mpc$^{-1}$, $\Omega_M = 0.3$, $\Omega_\Lambda = 0.7$.
 
\section{OBSERVATIONS AND SAMPLE SELECTION}

The IRAC data come from the IRAC Shallow Survey \citep{eis04}, which
is one of the major IRAC GTO programs.  The data cover 8.5~square
degrees in the Bo\"otes field of the NOAO Deep-Wide Field Survey
\citep[NDWFS][]{Jannuzi1999}.  Each sky location was observed with
three 30~s frames at 3.6, 4.5, 5.8, and 8.0~\micron. To create the
images in each waveband, the Basic Calibrated Data (BCD) were
mosaiced using the Spitzer Science Center software MOPEX
\citep{mor05}.  An 8~\micron\ catalog was created using SExtractor
\citep{ber96}.  The 5$\sigma$ limiting magnitude at 8.0~\micron\ is
14.9 in the Vega system, equivalent to a flux density of 69~$\mu$Jy
\citep{bro06}.  The galaxy surface density at this limit is about
10600 galaxies deg$^{-2}$ \citep{Fazio2004}.

Photometry in all four IRAC bands was performed using the SExtractor
double-image mode.  For each source in the 8~\micron\ catalog, two
magnitudes were derived \citep{bro06}: (1) an aperture magnitude
within a 5\arcsec\ diameter, corrected to total magnitude using the
point spread function (PSF) growth curve for point sources, and (2) a
SExtractor ``auto'' magnitude.  Figure~\ref{f:mags} shows the
difference between these two magnitudes as a function of the aperture
magnitude. For objects with angular diameters larger than a few
arcseconds, the aperture correction does not fully account for flux
outside the aperture, and the aperture magnitude therefore
underestimates the source flux.  For faint point sources and also for
extended sources with low surface brightness, SExtractor chooses too
small a size for its ``auto'' aperture, and the auto magnitudes
underestimate the flux \citep[e.g., Fig.~5
of][]{Labbe2005,Brown2007}.  In order to account for these biases, we
used the brighter of the two magnitude measurements.  For the largest
galaxies the auto aperture encircles almost all of the flux and
therefore gives total magnitude directly, while for faint galaxies,
which are nearly pointlike, the aperture magnitude calibrated for
point sources likewise lead to a total magnitude.  Flux densities for
low surface brightness galaxies may still be underestimated, but
these are a small fraction of the sample.\footnote{Less than 5\% of
galaxies have surface brightness lower than the equivalent of
13.5~mag in a 5\arcsec\ diameter.  This is a strong upper limit on
the fraction of galaxies that could be affected.}  All magnitudes in
this paper are reported in the Vega magnitude system.

In addition to the photometric data, the AGN and Galaxy Evolution
Survey (AGES, \citealt{koc06}) has measured 15,052 redshifts in the
Bo\"otes field with the Hectospec multi-object spectrograph
\citep{Fabricant1998, Roll1998, Fabricant2005} on the MMT.  Redshifts
have been measured for 92\% of galaxies to a magnitude limit of
$[8.0]=13.2$ and 65\% of galaxies to $[8.0]=13.8$ in a 6.88~deg$^2$
sub-field indicated in Figure~\ref{f:field}.  We limited the sample
to this subfield and to $[8.0]<13.5$ (251~$\mu$Jy) to minimize the
redshift incompleteness and to $z<0.3$ to minimize uncertainties in
the K-corrections.\footnote{The magnitude limits for the AGES
spectroscopy, $15\le I \le 20$, had no significant effect on the
sample selection.}  Beyond $z=0.3$, the strongest PAH feature at
7.7~\micron\ shifts out of the IRAC 8.0~\micron\ band.  The sample
was also limited to $z>0.02$ to avoid statistical noise from small
numbers of nearby galaxies.
Figure~\ref{f:redshift} shows the redshift completeness in the final
sample.  The sample limit corresponds to a limiting 8~\micron\
absolute magnitude of $-27.7$ at $z=0.3$ (neglecting the K-correction).

The 8.0~\micron-selected sample includes normal galaxies, AGN, and
stars. Figure~\ref{f:stern} shows the $[5.8]-[8.0]$ color as a
function of $[3.6]-[4.5]$ for all of the objects in the sample.  At
low redshift, AGN
and galaxies with PAH emission both have red IRAC colors, but the two
classes can be separated using IRAC color criteria
\citep{lac04,ste05}.  Galactic stars and elliptical galaxies are
mixed together at colors near (0,0); morphology was needed to
distinguish between them.  This sample is nearby enough that
elliptical galaxies are detectably extended in the NDWFS data. The
number of objects of each type are listed in Table~1 --- 964 objects,
14\% of the photometric sample, are identified as AGN.\footnote{
Some galaxies identified as ``normal'' probably have weak or obscured
AGN, but these AGN are not detectable in either the IRAC colors or the AGES
spectra and are therefore  contribute little if anything to the
observed light.} 
However, only 42 of the 549 AGN with a measured redshift are at
$z<0.3$ whereas 2556 out of the 3667 normal galaxies are at $z<0.3$.
The AGN fraction at $z<0.3$ is therefore about 1.6\% before allowing
for incompleteness.  Because AGES preferentially observed AGN
candidates, the true AGN fraction is likely to be even lower than
this.  Known AGN were excluded from the sample, but separate results
for them are explicitly mentioned in a few places below. As the
results show, unknown AGN should have an insignificant effect on the
derived luminosity functions.

\section{K-CORRECTIONS}

K-corrections for the galaxies in this sample are based on a simple
two-component model for the SEDs. In this model, the SED for every
galaxy is a linear combination of two fixed SEDs: (1) an old
``early-type'' stellar 
population, as might be found in an elliptical galaxy or spiral
bulge, and (2) a mix of stars and interstellar emission as might be
found in a ``late-type'' spiral galaxy disk.  For each of these
components, the SED 
was determined from an average of the observed SEDs for nearby
galaxies that have both 2MASS
photometry and ISO spectroscopy \citep{lu03}.  Ellipticals were used
to define the early-type SED, and disk galaxies were used to
define the late-type SED.\footnote{
  We use ``early-type'' and ``late-type'' as convenient terms to refer to the
  two SED components despite having no direct knowledge of the actual
  morphology of individual galaxies.  For example, irregular galaxies
  may have a ``late-type SED'' but not disk morphology.  Mid-infrared
  colors for the SED components are the opposite of those in visible
  light: late types are red, and early types are blue.}
The SED for each of these components was normalized at the rest-frame
$K$~band. Eleven model SEDs, ranging from a pure early-type template to a pure
late-type template,
were created. The late/early ratio for these model SEDs is defined as
the ratio of the flux densities of the late-type and early-type components in
the rest-frame $K$~band. In addition, a twelfth template matching the M82
SED was included to represent galaxies with high
star-formation rates.\footnote{M82 has an IR luminosity of
  $4\times10^{10}$~\Lsol\ \citep{Bell2003} corresponding to a SFR of
  about 7~\Msun~yr$^{-1}$.  Its SED is similar to
  SEDs of galaxies of even higher luminosity and SFR.}
Figure~\ref{f:templates} shows the templates from which the model
SEDs were constructed.
Convolving each model with the IRAC filter functions gave the
K-correction for each of the four IRAC bands as a function of redshift.
Figure~\ref{f:k} shows the results.  For $z<0.3$, the 3.6 and 4.5~\micron\
K-corrections are almost independent of the models, and the
8.0~\micron\ correction is also nearly model-independent  unless the galaxy is
early-type-dominated.  In contrast, the 5.8~\micron\
K-correction is highly variable because it depends critically on the
strength of  the 6.2~\micron\ PAH feature.

The suite of SED models spans the observed range of galaxy colors
with very few outliers and almost none at $z<0.3$.
Figure~\ref{f:cm8} shows the observed $[4.5]-[8.0]$ galaxy colors as
a function of redshift.  Given the observed color and redshift for
each galaxy, the proportion of early- and late-type SED components
was determined by interpolation.  This proportion defines the
rest-frame colors for the galaxy and thus the K-correction for each
of the IRAC magnitudes.  The few galaxies with $[4.5]-[8.0]$ colors
redder than M82 were assigned the M82 K-corrections.  AGN (when
considered at all) were assigned K-corrections based on a power law
SED between 5.8 and 8.0~\micron.

\section{RESULTS}
\subsection{Mid-Infrared Colors of the Sample Galaxies}

The galaxy sample divides into two populations in a color-magnitude
diagram.  Figure~\ref{f:cm8} shows  the 8.0~\micron\ absolute magnitude
as a function of rest-frame $[4.5]-[8.0]$ color. Galaxies in the red,
late-type population ($[4.5]-[8.0]_{\rm rest} > 1.1$, cf.\
Fig.~\ref{f:color}) are star-forming systems with prominent
8~\micron\ PAH emission.  According to \citet{pah04b}, this color
corresponds to Sa and later morphological types.  The blue, early-type
population consists of galaxies with weak or no PAH
emission corresponding to  S0/a and earlier types \citep{pah04b}.
Figure~\ref{f:cm36} shows a different color-magnitude diagram, this
one based on 3.6~\micron\ magnitude, which is a measure of the stellar
luminosity.  The early-type population generally has higher stellar
luminosity than the late-type population, and galaxies fainter than
$[3.6]_{\rm abs}=-24$ are uncommon in the early-type population but common
in the late-type
one.  Within the late-type population, the lack of correlation between the
rest-frame $[4.5]-[8.0]$ color and $[3.6]_{\rm abs}$ implies that the
3.6~\micron\ luminosity is little affected by the 3.3~\micron\ PAH
emission feature. In contrast, $[8.0]_{\rm abs}$ is strongly
correlated with the rest-frame $[3.6]-[4.5]$ color (Fig.~\ref{f:cm8short}),
especially for $[8.0]_{\rm abs}<-27$. This correlation suggests that
dust emission at 4.5~\micron\ is detectable in the star-forming late-type
population, \ie, galaxies with strong PAH emission also show dust
continuum emission at 4.5~\micron. This is consistent with ISOPHOT
spectroscopy: \citet{lu03} detected 4~\micron\ dust emission from
nearby spiral galaxies and suggested it arises from a dust component
closely related to the PAH.  The same UV sources that heat PAH molecules
may also heat the dust that produces the excess 4.5~\micron\
emission.  Dust emission at 4.5~\micron\ is relatively unimportant
for the least luminous galaxies, those with $[8.0]_{\rm abs}>-26$,
despite the presence of 8~\micron\ emission in a subset of the
low-luminosity galaxies (Fig.~\ref{f:cm8}).

The 8~\micron\ number counts are strongly dominated by the late-type
population: only 3\% of the spectroscopic sample have early-type colors.
Even for the early-type-dominated galaxies, most of the
8~\micron\ emission comes from dust rather than stellar light
(Fig.~\ref{f:color}).  The late-type galaxies are even more dominant in
8~\micron\ luminosity, as shown in Fig.~\ref{f:cm8}. All  the early-type
galaxies have 8~\micron\ absolute magnitudes at least three magnitudes fainter
than the brightest late-type galaxies.

\subsection{The 8.0~\micron\ Luminosity Function}

There are many ways to estimate the galaxy luminosity function from
survey data.  When examining a new wavelength regime, a
non-parametric method --- one not assuming a specific functional form
--- is desirable because the true luminosity function may not follow
a Schechter or other typical distribution.  For the present survey,
where galaxy distances vary by a factor of ten, it is also better to
avoid methods that are sensitive to possible galaxy
clustering. \citet{Willmer1997} and \citet{Takeuchi2000} have
presented detailed comparisons of many methods. We have computed the
luminosity functions with both the familiar $1/V_{\rm max}$ method
\citep{Schmidt1968}
and the \citet{LB1971} C$^-$~method.  The C$^-$~method does not
require binning the data, thus avoiding the possible bias in
$1/V_{\rm max}$ found by \citet{Page2000}, and it is insensitive to
possible galaxy clustering.  The results of the two methods are
consistent.  A further verification that large scale structure has
little effect was to ignore data in narrow but heavily populated
redshift ranges such as $z\approx0.13$ and $z\approx0.19$
(Fig.~\ref{f:color}).  The resulting luminosity functions do not
change significantly.

The luminosity functions have to be corrected for spectroscopic
incompleteness in the sample.  Because this sample has spectra for
nearly all galaxies (Table~1, Fig.~\ref{f:redshift}), the
incompleteness correction has an insignificant effect on the results
except in the faintest luminosity bins.\footnote{Incompleteness 
for bright {\em apparent} magnitudes --- mostly galaxies at small
distances --- is distributed across all luminosity bins and has
little effect on most of them because it affects only small numbers
of galaxies in each bin.}
This was checked by
recomputing the luminosity function with two different incompleteness
corrections.  One method was to use only galaxies with spectroscopic
redshifts but apply a weight to each one according to the reciprocal
probability it would have been observed in the spectroscopic sample.
The second method included all galaxies.  Galaxies were divided into
bins based on $[4.5]-[8.0]$ colors and $[8.0]$ magnitudes, and
galaxies without redshifts were assigned the same redshift
distribution as the galaxies with known redshifts in the same
bin.\footnote{In practice, this was achieved by weighting the
galaxies with known redshifts according to (total number)/(number
with known redshift) where the ratio was calculated  for
each bin.}  The two methods give luminosity functions that are equal
within the Poisson uncertainties.

The non-parametric 8.0~\micron\ luminosity functions for various
galaxy classes are shown in Figure~\ref{f:lfdiff} and Tables~2 and~3.  As
shown in this figure and also Figure~\ref{f:cm8}, blue
galaxies contribute little to the luminosity function and are rare
for $[8.0]_{\rm abs}<-26$.  AGN are insignificant except in
the brightest luminosity bins, where the uncertainties are large
because few objects are in the survey.  The luminosity function in
the range $-29.5 \la [8.0]_{\rm abs} \la -22$ is dominated by
galaxies with significant dust contribution to the 8~\micron\ flux.

In addition to total luminosity, the observations allow us to measure
the 8.0~\micron\ PAH (dust) luminosity alone.  For each galaxy in the
sample, the rest-frame 3.6~\micron\ flux density was scaled by a
factor of 0.227 \citep[an estimate of the stellar flux density at
8~\micron\ ---][]{pah04a} and subtracted from the rest-frame 8.0~\micron\
flux density.  This subtracts the starlight, leaving the PAH flux
density. Because the subtraction is done on {\em rest-frame} flux
densities, the procedure is equivalent to the two-component
decomposition illustrated in Fig.~\ref{f:color}.  After subtracting
the continuum, galaxies with residual 8.0~\micron\ magnitudes fainter
than the limiting magnitude $[8.0]=13.5$ were excluded from the
sample, and the luminosity function was re-calculated for the
remaining galaxies to give a PAH-only luminosity function.  The
result is shown in Figure~\ref{f:lfdiff} and Tables~2 and~3.

The uncertainties in the luminosity functions are set both by Poisson
statistics of the sample and by cosmic variance.  The Poisson
uncertainties for the $1/V_{\rm max}$ method are given in Table~2 and
statistical uncertainties for 
the $C^-$ method in Table~3.  Cosmic variance uncertainty can
be estimated \citep{dh} from the volume sampled plus an estimate of
the galaxy two-point correlation function.  This is not known
directly for the 8~\micron\ sample, but we assume it's the same as
for optically-selected galaxy samples.  (If anything, the correlation
function is likely to be smaller for 8~\micron-selected galaxies,
which are predominantly late-type, than for optically-selected ones,
more of which are early-type --- \citealt{Norberg2002}.  This will
lead to less cosmic variance for an 8~\micron-selected sample.)  In
practice, a correlation function based on the fluctuation power
spectrum from WMAP-1 \citep{Spergel2003} was extrapolated to the
present via the method of \citet{Seljak1996} then transformed via a
spherical Bessel function and smoothed with a 1~Mpc radius.  The
volume integral \citep[e.g.,][Eq.~1]{Newman2002} was then evaluated
via a Monte Carlo approach.  The resulting cosmic variance
uncertainty is $\approx$15\% in bins $[8.0]_{\rm abs}<-27.7$, where
the survey samples the full volume of $z\le0.3$.  This is a single
uncertainty for the entire luminosity function, not an independent
uncertainty in each bin.  In the fainter bins, the uncertainty should
in principle be larger, but the entire method is based on the
assumption that the luminosity function is constant within the volume
surveyed.  In the faintest bin, for example, the survey magnitude
limit corresponds to 135~Mpc, and the data give no information
whatsoever on the luminosity function outside this distance.  Density
peaks in the sample are evident in Figure~\ref{f:color} but do not
affect the derived luminosity function.

As it turns out, a Schechter function is an excellent fit to both the
total 8.0~\micron\ and the PAH luminosity functions, as shown in
Figure~\ref{f:lfdiff}.  Table~4 gives the best-fit parameters as
found using the STY maximum likelihood method \citep{sty79}, and
Figure~\ref{f:ellipse} shows the interdependence of the derived
characteristic magnitudes $M^*$\footnote{The characteristic absolute magnitude
values refer to luminosity $\nu L_\nu$
emerging from the galaxy near 8~\micron\
measured in units of the Sun's bolometric luminosity.}
and faint-end slopes $\alpha$.    
The derived Schechter exponents for the middle to faint
end slopes are $\alpha = -1.38$ and $\alpha = -1.26$
respectively.   The slopes are similar to those
found in the Sloan 
$r$~band \citep{Loveday2004,Xia2006} and the infrared $K$ band
\citep{Huang2003,Jones2006} over a similar redshift range, though
\citet{Blanton2003} found a slightly shallower slope in $r$.
The steeper slope  for the total
luminosity shows that fainter galaxies tend to have relative less PAH
emission compared to their stellar emission  (though the slope
uncertainties overlap at the 2$\sigma$ level, as seen in 
Fig.~\ref{f:ellipse} and Table~4).  The inactive ``blue''
galaxies have just as steep a faint-end slope ($\alpha=-1.39$) as the
entire sample, in contrast to the result in blue light, where
inactive galaxies show a much shallower 
faint-end slope \citep{Madgwick2002}. At the bright end,
$L>L^*$, there are no excess galaxy numbers above the Schechter
function prediction, in contrast to the results of \citet{sau90} for
the IRAS 60~\micron\ luminosity function.  The high luminosity
population, \ie, LIRGs and ULIRGs, consists of galaxies with
far-infrared color temperatures $>$36~K \citep{sau90}, characteristic
of Seyfert and starburst galaxies.  Our exclusion of AGN in the
8.0~\micron\ sample, which removed galaxies from the bright end of
the luminosity function, does not explain why a Schechter function
can fit the remainder of the sample.

Integrating the luminosity functions over the whole sample gives the
total 8~\micron\ luminosity density in the local Universe.  The
result is $3.1 \times 10^7~$\Lsun~Mpc$^{-3}$ for total 8~\micron\
luminosity density (not including AGN), $2.5 \times
10^7~\Lsun$~Mpc$^{-3}$ for PAH luminosity density, and a highly
uncertain $0.05 \times 10^7~\Lsun$~Mpc$^{-3}$ for AGN luminosity
density.  (The sample is too small to determine reliable AGN
numbers.)  Thus over the entire galaxy population, stellar emission
contributes about 19\% of the 8.0~\micron\ luminosity while dust
emission contributes 80\% and AGN roughly 1\%.  Galaxies with 
$2\times 10^9~\Lsun <\nu L_{\nu}(8.0\micron) < 2\times10^{10}$~\Lsun\
contribute 50\% of the total 8~\micron\ luminosity, and galaxies
above and below this range contribute about 25\% each.
Thus the dominant contribution to the PAH luminosity density comes from
normal galaxies, not LIRGs or ULIRGs.
\label{s:pah} 

\subsection{Star Formation Rates}

Star formation rates in the local universe have been measured by a
variety of techniques.  (See the review by \citealt{Kennicutt1998a}.)
\citet{hop04} compiled star formation rate densities (SFRD) from a
variety of tracers.  According to Hopkins' compilation, over the
$z<0.3$ range of the present survey, the volume-weighted 
average ${\rm SFRD} \approx 0.031$~\Msun~yr$^{-1}$~Mpc$^{-3}$, though
some tracers tend to be systematically higher (emission lines) or lower
(ultraviolet) than this value.
\citet{Martin2005} combined ultraviolet with
far infrared data and found a local ($z<0.04$) ${\rm SFRD} \approx
0.021$~\Msun~yr$^{-1}$~Mpc$^{-3}$, about 20\% higher than but
consistent with Hopkins' result at this much smaller redshift.

Dust emission at 8.0~\micron\ is correlated with the total infrared
luminosity $L_{IR}$ in galaxies \citep{rig99,elb02,lu03} although the
relation is non-linear \citep{lu03} and has large scatter \citep[\eg,
Fig.~17 of][]{smith07}.  In normal star-forming galaxies, the
integrated flux between 5.8 and 11.3~\micron\ is between 9\% (for
galaxies \citeauthor{lu03} call ``FIR-quiescent'') and 5\% (for
\citeauthor{lu03} ``FIR-active'' galaxies) of $L_{IR}$ \citep{lu03}.
Recent results from \s\ are consistent with these values;
\citet{smith07} find that in the central areas of normal galaxies,
the 7.7~\micron\ PAH complex can contribute up to 10\% of $L_{IR}$.
In ULIRGs, PAH emission in the 5.8--11.3~\micron\ range is lower at
$\la$1\% of $L_{IR}$ 
\citep{rig99,Armus2007}; this could be partly due to an AGN contributing to
$L_{IR}$. Despite the uncertainties, some authors 
\citep[e.g.,][]{FS2004,Wu2005} have suggested that $\nu L_\nu ({\rm
PAH})$ can be used to measure star formation rates for galaxies, and
\citet{Wu2005} derived ${\rm SFR} \approx \nu L_\nu ({\rm PAH}) /
(1.5 \times 10^9~\Lsun)$~\Msun~yr$^{-1}$. This result was based on a
sample of Sloan galaxies showing H$\alpha$ emission.  The PAH
luminosity density found in \S\ref{s:pah} combined with the
local SFRD \citep[Fig.~2]{hop04} integrated from $z=0$ to 0.3
gives a volume average ${\rm SFR} \approx
\nu L_\nu ({\rm PAH}) / (0.84 \times 10^9~\Lsun)$~\Msun~yr$^{-1}$,
about double that found by \citeauthor{Wu2005} for individual
galaxies.  In other words, the volume-averaged PAH luminosity density
is half as much as expected, given the volume-averaged
SFRD and ${\rm SFRD} / \nu L_\nu({\rm PAH})$ ratios of individual galaxies.
The difference suggests that the galaxy samples have been weighted to
galaxies with unusually large PAH emission for their SFRs, but this
is hard to understand if the dominant contribution to star formation
comes from the normal galaxy population \citep[e.g.,][]{flo05}.

\section{SUMMARY}
  
An 8.0~\micron-selected sample of low redshift ($z<0.3$) galaxies
finds primarily star-forming galaxies.  AGN represent less than 2\%
of the sample both by number and by fraction of the
8~\micron\ luminosity.  The luminosity function of the remaining
galaxies can be described as a Schechter function with
characteristic magnitude $M^*=-28.46$ 
(corresponding to $\nu L_{\nu} = 1.8 \times
10^{10}$~\Lsun) and $\alpha=-1.38$.
Subtracting starlight continuum emission from each galaxy gives a
luminosity function for emission from the interstellar
medium component (mostly PAH) alone.  This can also be approximated
by a Schechter 
function with characteristic magnitude $M^*=-28.34$ ($\nu L_{\nu} = 1.6
\times 10^{10}$~\Lsun) and $\alpha=-1.26$.

The total 8.0~\micron\ luminosity density for this 8.0~\micron-selected
sample is $3.1 \times 10^7$~\Lsun~Mpc$^{-3}$. Ignoring the luminosity
contributed by AGN, about 81\% of the 8~\micron\ luminosity is from
interstellar dust emission, presumably PAH, and about 19\% is from
stellar photospheres.  The observed, volume-averaged ratio of PAH
luminosity density to star formation rate density is about half of
an estimate based on local, individual galaxies.

\acknowledgements

This work is based in part on observations made with the Spitzer
Space Telescope, which is operated by the Jet Propulsion Laboratory,
California Institute of Technology under a contract with
NASA. Support for this work was provided by NASA through an award
issued by JPL/Caltech.  JLR has received support from an NSF
Astronomy and Astrophysics Postdoctoral Fellowship under grant
AST-0302049.  Observations reported here were obtained at the MMT
Observatory, a joint facility of the Smithsonian Institution and the
University of Arizona.  This work made use of images and/or data
products provided by the NOAO Deep Wide-Field Survey (Jannuzi and Dey
1999), which is supported by the National Optical Astronomy
Observatory (NOAO). NOAO is operated by AURA, Inc., under a
cooperative agreement with the National Science Foundation.

Facilities: \facility{Spitzer(IRAC),MMT(Hectospec)}





\clearpage
\begin{figure}
%\plotone{f1.ps}
\caption{Difference between the SExtracter total (``auto'')
magnitudes and 5\arcsec\ aperture magnitudes.  Offsets are mostly
positive, indicating that aperture magnitudes underestimate the total
flux of resolved galaxies. We selected the sample using the aperture
magnitudes to avoid bias against lower surface brightness galaxies,
and set the magnitude limit at $[8.0]=13.5$. The derived luminosity
functions are based on the auto magnitudes for the resolved galaxies
and the aperture magnitudes for the unresolved galaxies.
\label{f:mags}}
\end{figure}

\clearpage
\begin{figure}
%

%\plotone{f2.ps}
\caption{The B\"ootes field showing the approximate area of 8~\micron\
  coverage outlined in red. Positions are J2000. Black dots denote
  denote the 8.0~\micron-detected
  galaxies that have AGES spectroscopic redshifts. The black lines enclose
  the area of the highly complete redshift sample used for this paper.
\label{f:field}}
\end{figure}

\clearpage
\begin{figure}
\plotone{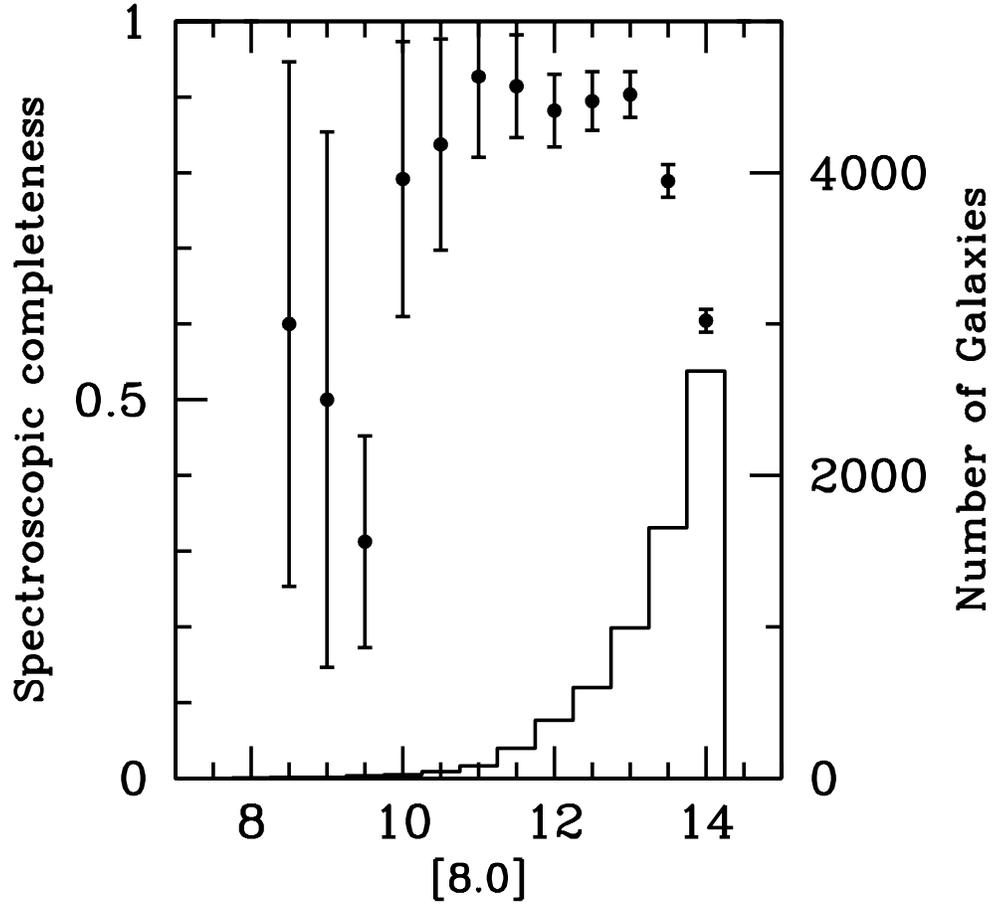}
\caption{Redshift identification fraction in each
half-magnitude bin (filled circles, left ordinate label). Error bars
show the Poisson 
uncertainty.  The solid line shows the histogram of the 8~\micron\
magnitude distribution of all galaxies in the catalog (right ordinate
label). Only galaxies with $[8.0]<13.5$ are in the final sample.
\label{f:redshift}}
\end{figure}

\clearpage
\begin{figure}
\begin{center}
\end{center}
\caption{$[5.8]-[8.0]$ vs $[3.6]-[4.5]$ color-color diagram for
separation of AGN, star-forming galaxies, and early type galaxies
\citep{ste05}. Data are plotted as observed, \ie, before
K-correction.  Black dots denote normal galaxies, red dots denote
stars, and blue dots denote AGN, all as classified by optical
spectroscopy. The two thick lines show two SED models from $0<z<0.6$,
one with and the other without PAH emission. (The latter is almost
hidden behind the data near zero color.)  The thin solid lines show
the \citet{ste05} AGN-galaxy separation criterion; AGN lie within the
quadrilateral extending to the top of the plot. The color cut at
$[5.8]-[8.0]=0.5$ 
separates stars and elliptical galaxies from galaxies with
significant PAH emission.  Galaxies with PAH emission cannot enter
the ``no PAH'' area unless their redshifts are $>$0.6.
\label{f:stern}}
\end{figure}

\clearpage
\begin{figure}
\plotone{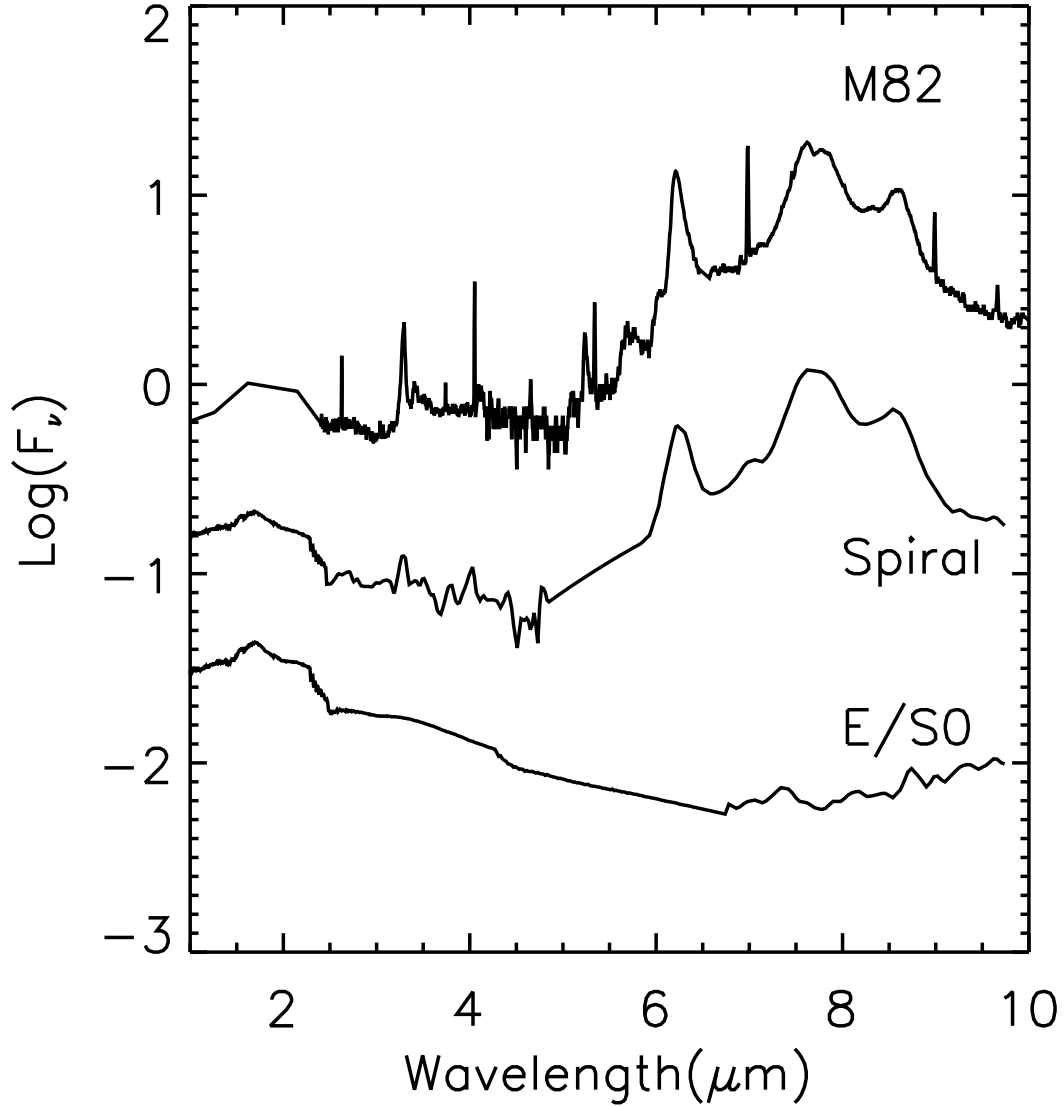}
\caption{Galaxy SED templates.  The upper curve shows the M82
template, the middle curve the spiral (``late-type'') template, and
the bottom curve the elliptical (``early-type'') template.  The
vertical axis is in arbitrary units.  The early-type template is an
average ISOPHOT-S spectrum of six classical elliptical galaxies
compiled by \citet{lu00}.  It is very similar to the elliptical
spectrum shown by \citet{lu03} except at wavelengths longer than
8~\micron, where an improved ISOPHOT-S calibration made the current
spectrum slightly steeper (i.e., bluer).  The late-type template is
the average for spiral galaxies from \citet{lu00}.  Nine additional
templates used in this paper are mixtures of the early- and late-type
templates.
\label{f:templates}}
\end{figure}

\clearpage
\begin{figure}
\plotone{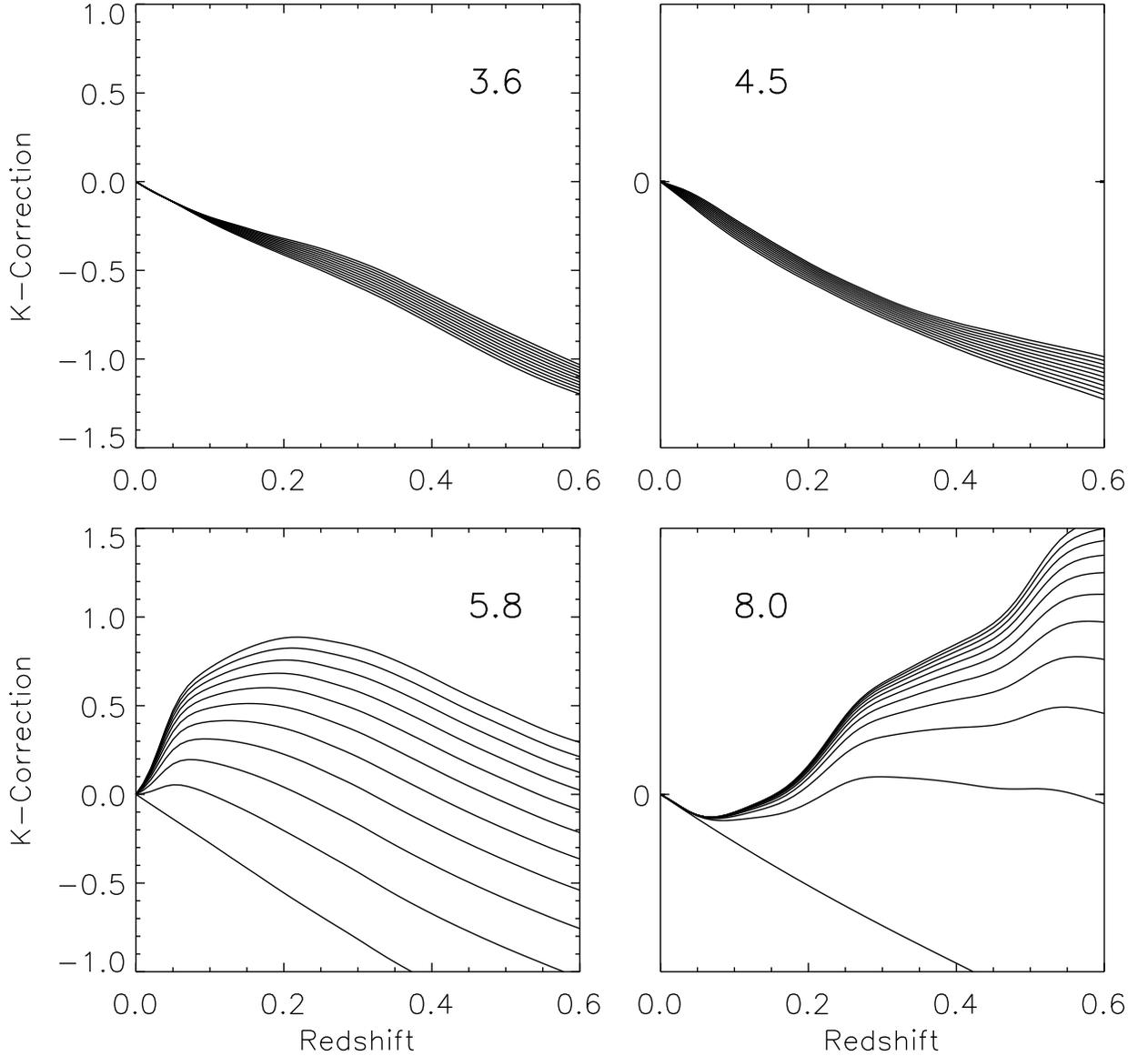}
\vspace*{-48pt}
\caption{K-corrections for the four IRAC channels for each of the
twelve SED models.  The bottom line in each panel shows the
early-type model.  The 3.6 and 4.5~\micron\ bands sample
mainly stellar emission longward of the Wien peak and therefore show
a negative K-correction.  The 5.8~\micron\ band is relatively narrow
and shows a positive K-correction, depending on the amount of PAH
present, as the 6.2~\micron\ PAH feature leaves the band at
$z\approx0.05$. As redshift increases further, the band samples only
stellar emission, and the K-correction decreases.  The 8~\micron\
band is wide, and the 7.7~\micron\ feature remains inside the band at
low redshift.  As redshift increases, the 6.2~\micron\ PAH feature
enters the band before the 7.7~\micron\ feature leaves, leading to a
nearly constant K-correction.  At $z\ga0.2$, the 7.7~\micron\ feature
begins to leave the band, and the K-correction increases according to
the PAH strength.
\label{f:k}}
\end{figure}

\clearpage
\begin{figure}
\plotone{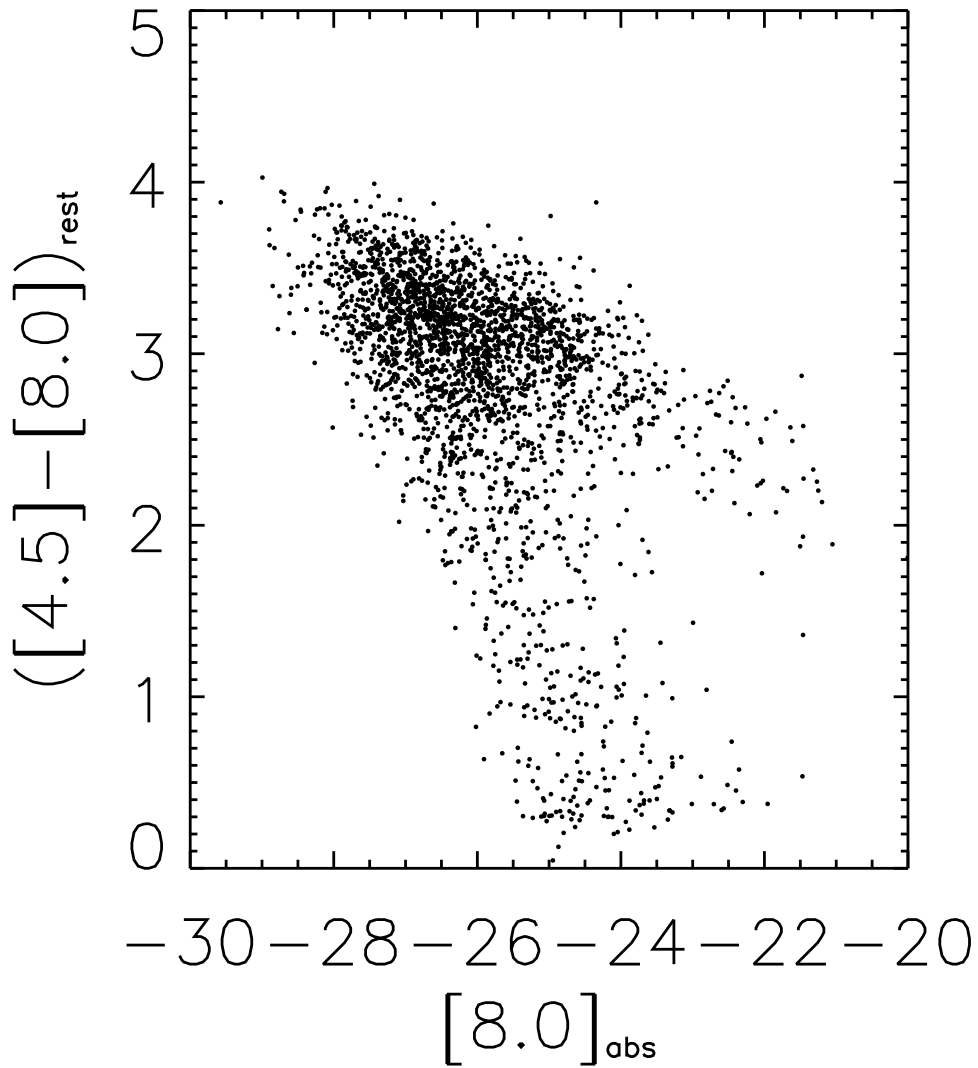}
\caption{
  Color-magnitude diagram showing $[4.5]-[8.0]$ color versus 8~\micron\
  absolute magnitudes.  Points denote galaxies with spectroscopic
  redshifts.  Colors and magnitudes are both corrected to the rest
  frame. Most galaxies show strong PAH emission ($[4.5]-[8.0]>2$) and
  are therefore in the ``late-type'' category as we have defined
  it. ``Early-type'' galaxies with weak or no PAH emission
  ($[4.5]-[8.0]\la0.4$) are well separated in this diagram from the
  much larger late-type population, but galaxies span the entire color range.
  None of the early-type galaxies has high luminosity at
  8~\micron.
\label{f:cm8}}
\end{figure}

\clearpage
\begin{figure}
\plotone{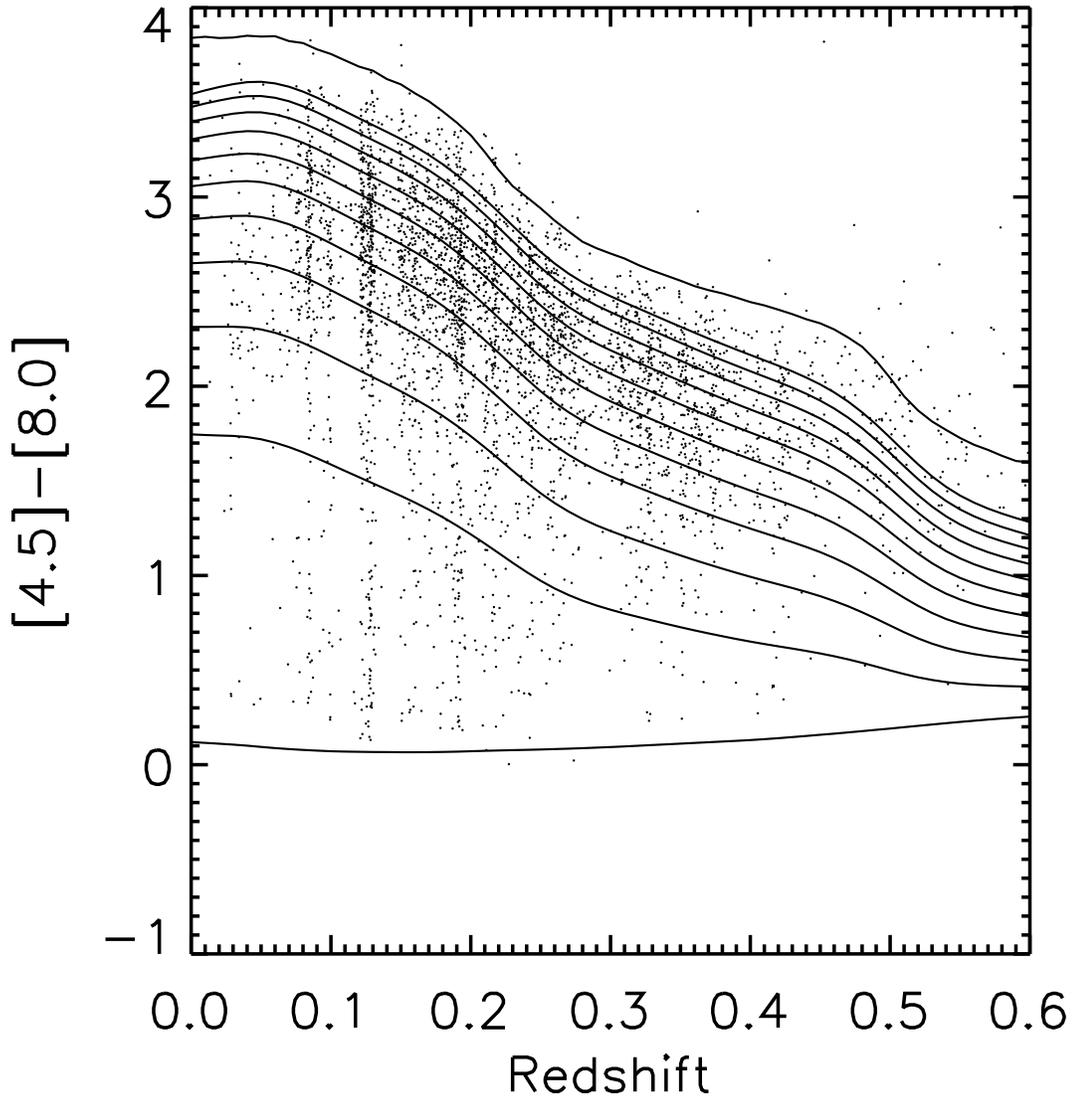}
\caption{\footnotesize The $[4.5]-[8.0]$ color versus redshift for
galaxies with spectroscopic redshifts. Points denote the observed
galaxies with AGN excluded.  Lines show colors for each of twelve
model SEDs as a function of redshift.  The reddest model is derived
from the M~82 SED. Other models are a linear combination of late-type
and early-type templates as shown in Fig.~\ref{f:templates}.  These
SEDs are normalized at rest 2.2~\micron\ with the 
stellar fraction going from 0 to 1 in steps of 0.1 from the bluest
model to the reddest one (second-reddest including the M82 model).
Only galaxies with $z<0.3$ are used in this paper.
\label{f:color}}
\end{figure}

\clearpage
\begin{figure}
\plotone{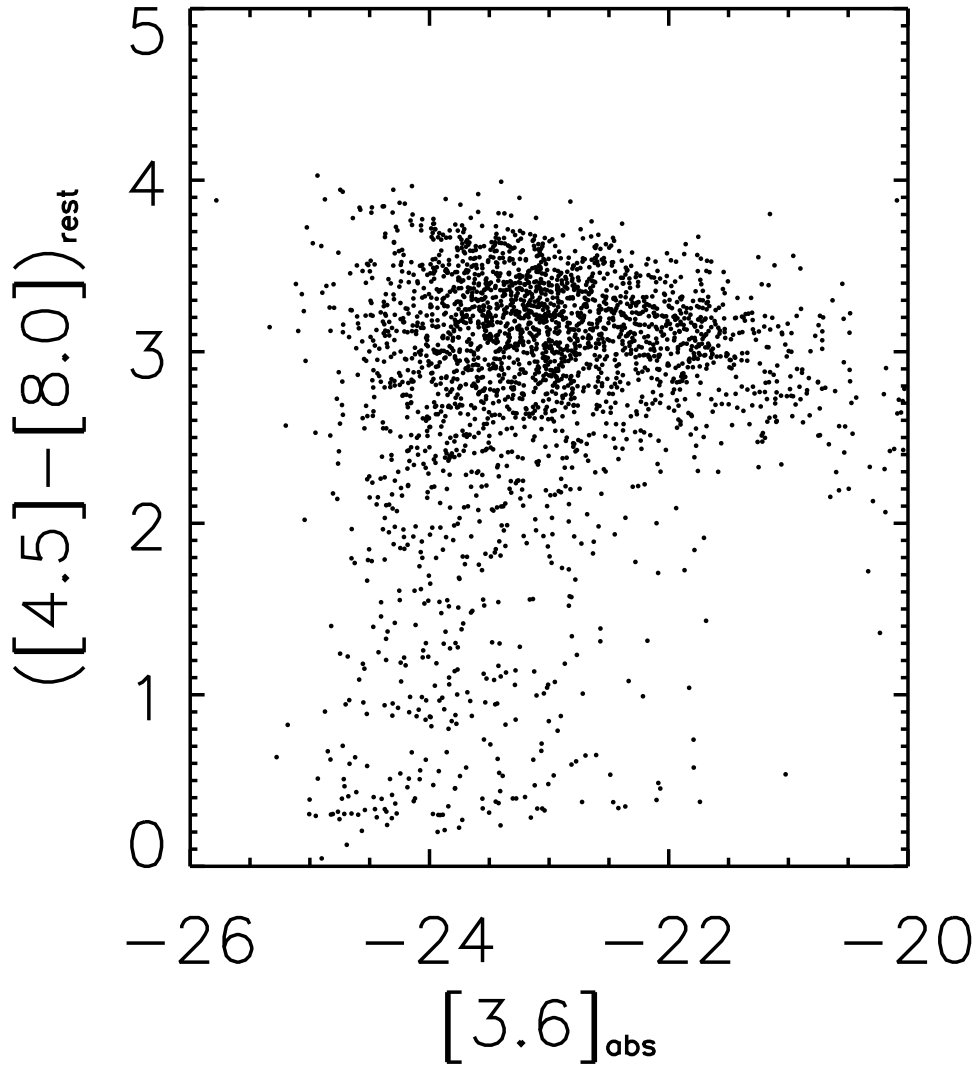}
\caption{Color-magnitude diagram showing $[4.5]-[8.0]$ color vs
  3.6~\micron\ absolute magnitudes.  Points denote galaxies with
  spectroscopic redshifts.  Colors and magnitudes are both corrected
  to the rest frame.  The upper luminosity envelope is populated by
  both blue ($[4.5]-[8.0]<1.1$) and star-forming galaxies, but there
  are no blue  galaxies with low 3.6~\micron\ luminosities.
  For the strongly star-forming galaxies, $[4.5]-[8.0] \ga 2$, color
  does not correlate with the absolute 3.6~\micron\ magnitude.
\label{f:cm36}}
\end{figure}

\clearpage
\begin{figure}
\plotone{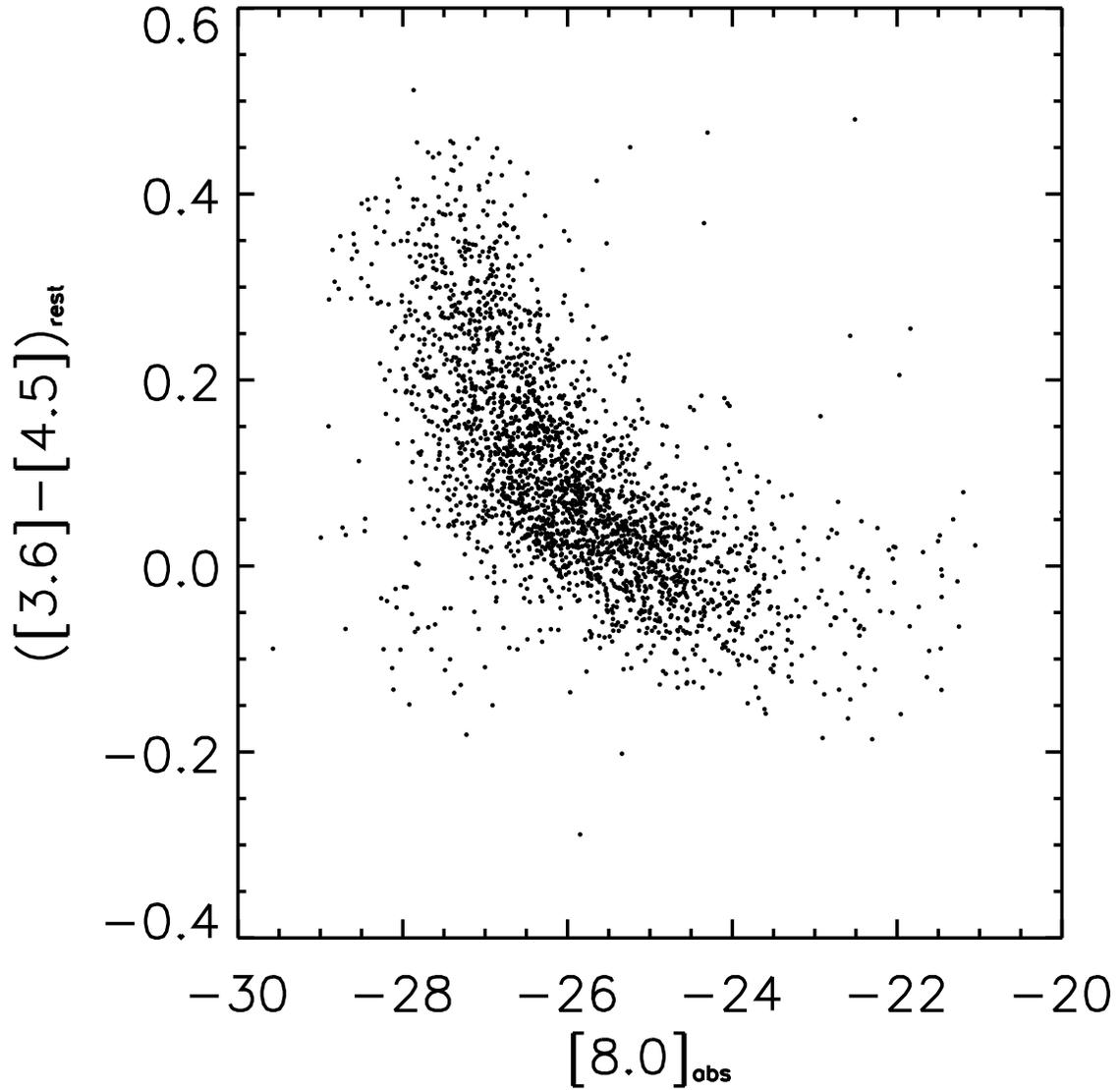}
\caption{Color-magnitude diagram showing $[3.6]-[4.5]$ color vs
  8~\micron\ absolute magnitudes.  Points denote galaxies with
  spectroscopic redshifts.  Colors and magnitudes are both corrected
  to the rest frame.  The correlation between color and magnitude
  indicates that dust emission may constitute about 30\% of the total
  4.5\micron\ emission for galaxies with $[8.0]_{\rm abs} \la -27$.
\label{f:cm8short}}
\end{figure}

\clearpage
\begin{figure}
\begin{center}
\includegraphics[height=5.5in, trim=108 108 0 54]{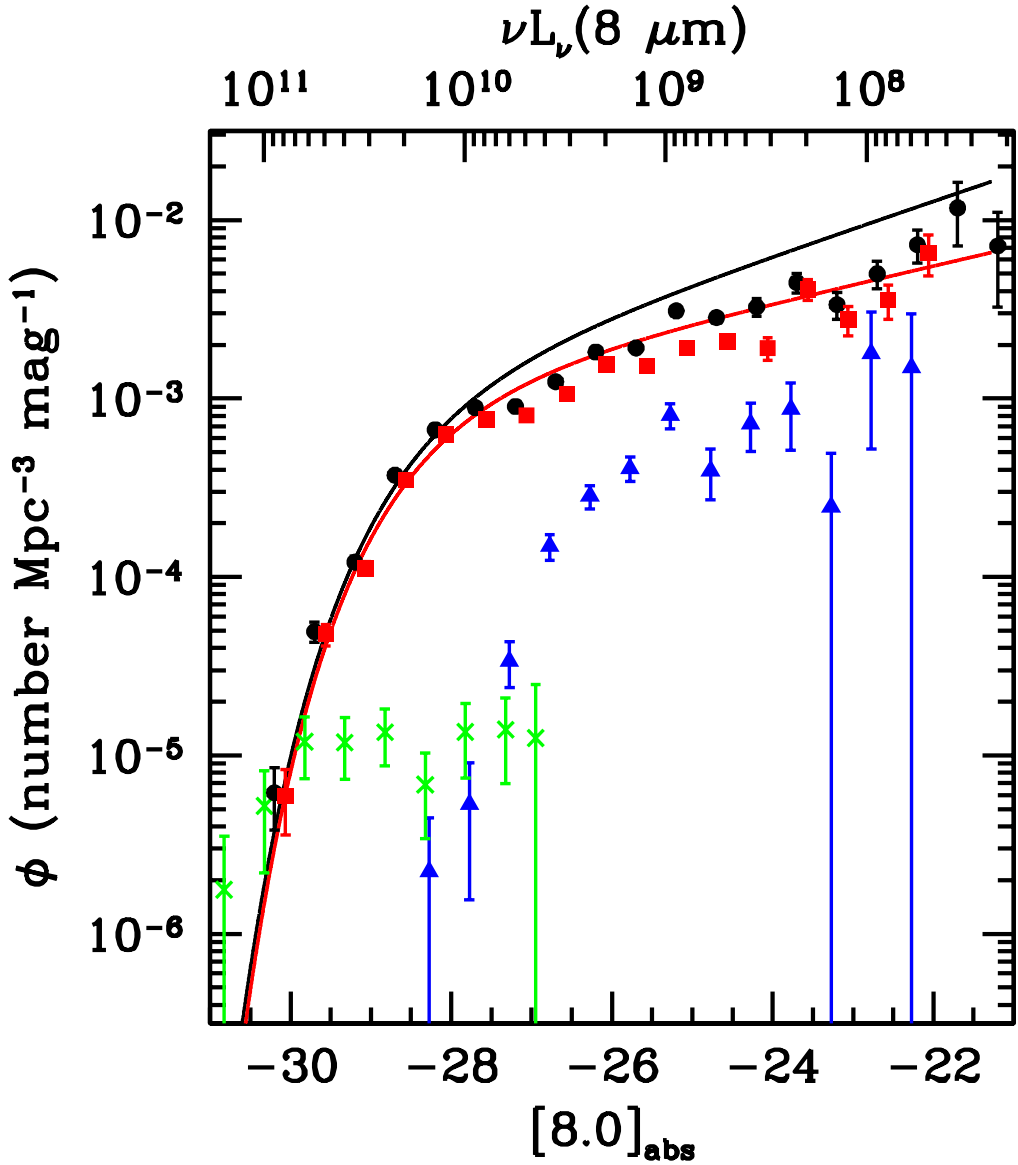}
\end{center}
\caption{\footnotesize
  Derived differential luminosity functions corrected for
  spectroscopic incompleteness. Black dots denote the 8~\micron\
  luminosity function for all galaxies excluding AGN, and red squares
  denote the luminosity function for PAH alone, \ie, after
  subtracting the stellar continuum emission for each galaxy.  These
  two luminosity functions were derived from the non-parametric C$^-$
  method.    Blue triangles
  denote the luminosity function 
  for early-type galaxies, specifically those with
  $[4.5]-[8.0]<1.1$.  Even for these galaxies, much of the emission
  is PAH despite the color selection.  Green crosses denote the
  luminosity function for AGN.  These latter two luminosity functions
  were derived from the $1/V_{max}$ method because of the smaller
  numbers of galaxies.  Because of the small numbers, the blue galaxy
  luminosity function is only reliable for $-27\la [8.0]_{abs} \la
  -25$.  The AGN luminosity function is highly uncertain throughout;
  only one bin has as many as 8 galaxies in it.  All points are
  slightly offset horizontally for clarity; see Tables~2 and~3 for actual
  magnitude bins.  Error bars show the Poisson uncertainties in each
  bin.  Lines show the respective parametric Schechter functions that best fit
  the data for ``all 8~\micron\ emission'' and ``PAH only.'' Both were derived
  from the STY method.  The bottom axis is labeled in 8~\micron\ Vega
  magnitudes, while the top axis is labeled in the equivalent $\nu
  L_\nu$ units.
\label{f:lfdiff}}
\end{figure}

\clearpage
\begin{figure}
\includegraphics[angle=-90]{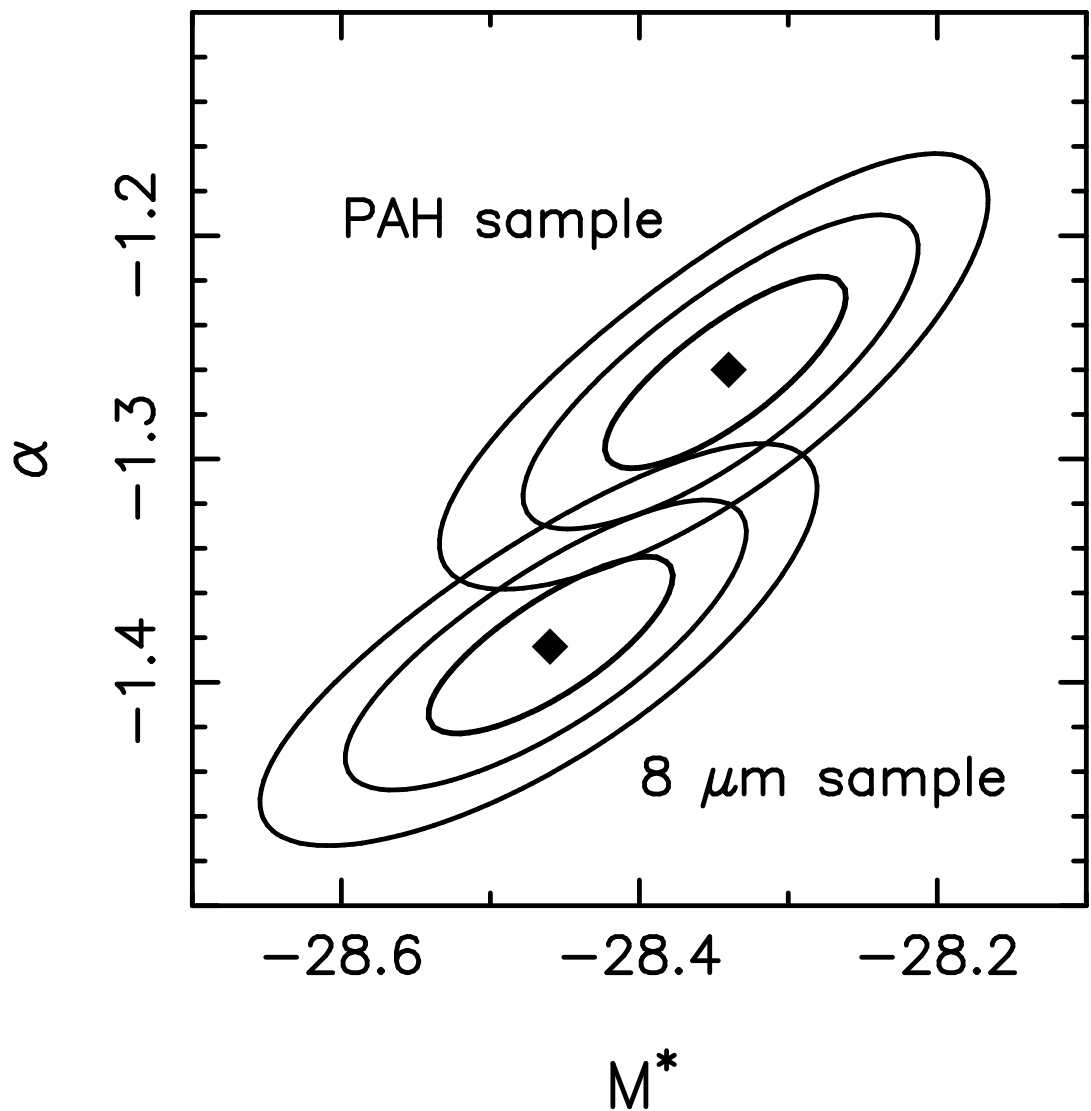}
\caption{Uncertainty ellipses for characteristic magnitude $M^*$ and
faint end slope $\alpha$ of the Schechter luminosity functions. The
ellipses were calculated using a grid in $M^*$--$\alpha$ space and
the STY formalism. The curves represent contours of constant $\Delta
\chi^2$ corresponding to 1$\sigma$ (inner), 2$\sigma$ (middle), and
3$\sigma$ (outer) confidence levels assuming three fitting parameters.
Statistical uncertainty in the overall normalization $\Phi^*$ is very
small, but the true uncertainty is set by cosmic variance (\S4.2).
Faint end slopes probably differ but are consistent at the 2$\sigma$ level.
\label{f:ellipse}}
\end{figure}


\clearpage

\begin{table*}[hbt]
{%\scriptsize
\begin{center}
\centerline{\sc Table 1}
\vspace{0.1cm}
\centerline{\sc Number of Objects at $[8.0]<13.5$}
\vspace{0.3cm}
\begin{tabular}{lccc}
\hline\hline
\noalign{\smallskip}
Object & Total  & Redshift Sample & $z<0.3$\cr
\hline
\noalign{\smallskip}
Star & 1894 & 0 & 0\cr
Blue Galaxies & 236 & 109 & 100\cr
Star-Forming Galaxies & 3667 & 3174 & 2556\cr
AGN & 964 & 549 & 42\cr
\noalign{\hrule}
\noalign{\smallskip}
\end{tabular}
\end{center}
}
\label{tab1}
\end{table*}

\clearpage
\begin{table*}[hbt]
{%\scriptsize
\begin{center}
\centerline{\sc Table 2}
\vspace{0.1cm}
\centerline{\sc Luminosity Functions Calculated by the $1/V_{max}$ Method }
\vspace{0.3cm}
\begin{tabular}{ccrccrcc}
\hline\hline
\noalign{\smallskip}
[8.0]$_{abs}$\tablenotemark{a}
& $\nu L_\nu(8~\micron)$
& $N$ 
& $\phi$(8.0\micron)
& $\Delta\phi$(8.0\micron)
& $N$
& $\phi$(PAH)
& $\Delta\phi$(PAH)\\

\hline
\noalign{\smallskip}
&&&\multicolumn{2}{c}{$10^{-3}$~Mpc$^{-3}$~mag$^{-1}$}&&
\multicolumn{2}{c}{$10^{-3}$~Mpc$^{-3}$~mag$^{-1}$}\\
-30.27& $9.5\times10^{10}$ &  2&\00.003  & \00.002&  2&\00.003  &\00.002 \\
-29.77& $6.0\times10^{10}$ & 10&\00.018  & \00.006&  9&\00.016  &\00.005   \\
-29.27& $3.8\times10^{10}$ & 37&\00.073  & \00.012& 31&\00.062  &\00.011  \\
-28.77& $2.4\times10^{10}$ &133&  0.25   & 0.02   &116& 0.22   & 0.02   \\
-28.27& $1.5\times10^{10} $&294&  0.56   & 0.03   &268& 0.50   & 0.03   \\
-27.77& $9.5\times10^{9\0}$&447&  1.06   & 0.05   &384& 0.86   & 0.04   \\
-27.27& $6.0\times10^{9\0}$&429&  1.45   & 0.07   &416& 1.34   & 0.07   \\
-26.77& $3.8\times10^{9\0}$&413&  2.07   & 0.10   &367& 1.75   & 0.09   \\
-26.27& $2.4\times10^{9\0}$&327&  2.49   & 0.14   &270& 1.94   & 0.12   \\
-25.77& $1.5\times10^{9\0}$&218&  2.71   & 0.18   &179& 2.04   & 0.15  \\
-25.27& $9.5\times10^{8\0}$&137&  3.2\0  & 0.3\0  & 98& 2.02   & 0.20  \\
-24.77& $6.0\times10^{8\0}$& 63&  2.5\0  & 0.3\0  & 64& 2.17   & 0.27  \\
-24.27& $3.8\times10^{8\0}$& 41&  3.0\0  & 0.5\0  & 27& 1.73   & 0.33  \\
-23.77& $2.4\times10^{8\0}$& 23&  3.7\0  & 0.8\0  & 13& 1.60   & 0.44    \\
-23.27& $1.5\times10^{8\0}$&  9&  2.4\0  & 0.8\0  & 16& 3.0\0  & 0.8\0  \\
-22.77& $9.5\times10^{7\0}$& 12&  7.4\0  & 2.1\0  &  7& 3.2\0  & 1.2\0  \\
-22.27& $6.0\times10^{7\0}$&  7&  7.7\0  & 2.9\0  &  6& 5.8\0  & 2.4\0  \\
-21.77& $3.8\times10^{7\0}$&  1&  2.4\0  & 2.4\0  &\no& \no    & \no    \\
-21.27& $2.4\times10^{7\0}$&  2& 10.1\0\0& 7.6\0  &\no& \no    & \no    \\
\noalign{\hrule}
\noalign{\smallskip}
\end{tabular}
\end{center}
}
\label{tab2}
\tablenotetext{a}
{Absolute Vega magnitude bin centers.}
\end{table*}

\clearpage
\begin{table*}[hbt]
{%\scriptsize
\begin{center}
\centerline{\sc Table 3}
\vspace{0.1cm}
\centerline{\sc Luminosity Functions Calculated by the C$^-$ Method }
\vspace{0.3cm}
\begin{tabular}{cccccc}
\hline\hline
\noalign{\smallskip}
[8.0]$_{abs}$\tablenotemark{a}
& $\nu L_\nu(8~\micron)$
& $\phi$(8.0\micron)
& $\Delta\phi$(8.0\micron)
& $\phi$(PAH)
& $\Delta\phi$(PAH)\\

\hline
\noalign{\smallskip}
&&\multicolumn{2}{c}{$10^{-3}$~Mpc$^{-3}$~mag$^{-1}$}&
\multicolumn{2}{c}{$10^{-3}$~Mpc$^{-3}$~mag$^{-1}$}\\
-30.13& $8.5\times10^{10}$ &\00.006  & \00.002&\00.006  &\00.002 \\
-29.63& $5.3\times10^{10}$ &\00.049  & \00.007&\00.048  &\00.007   \\
-29.13& $3.4\times10^{10}$ &\00.121  & \00.010&\00.112  &\00.009  \\
-28.63& $2.1\times10^{10}$ &  0.37   & 0.02   & 0.35   & 0.02   \\
-28.13& $1.5\times10^{10} $&  0.67   & 0.02   & 0.63   & 0.02   \\
-27.63& $8.5\times10^{9\0}$&  0.89   & 0.02   & 0.76   & 0.02   \\
-27.13& $5.3\times10^{9\0}$&  0.90   & 0.02   & 0.80   & 0.02   \\
-26.63& $3.4\times10^{9\0}$&  1.24   & 0.03   & 1.06   & 0.03   \\
-26.13& $2.1\times10^{9\0}$&  1.82   & 0.06   & 1.54   & 0.05   \\
-25.63& $1.5\times10^{9\0}$&  1.92   & 0.08   & 1.53   & 0.07  \\
-25.13& $8.5\times10^{8\0}$&  3.1\0  & 0.2\0  & 1.92   & 0.12  \\
-24.63& $5.3\times10^{8\0}$&  2.8\0  & 0.2\0  & 2.10   & 0.17  \\
-24.13& $3.4\times10^{8\0}$&  3.3\0  & 0.4\0  & 1.91   & 0.27  \\
-23.63& $2.1\times10^{8\0}$&  4.5\0  & 0.6\0  & 4.1\0  & 0.5\0   \\
-23.13& $1.5\times10^{8\0}$&  3.4\0  & 0.6\0  & 2.8\0  & 0.5\0  \\
-22.63& $8.5\times10^{7\0}$&  5.0\0  & 0.9\0  & 3.6\0  & 0.8\0  \\
-22.13& $5.3\times10^{7\0}$&  7.3\0  & 1.5\0  & 6.6\0  & 1.7\0  \\
-21.63& $3.4\times10^{7\0}$& 11.7\0\0& 4.5\0  & \no    & \no    \\
-21.13& $2.1\times10^{7\0}$&  7.2\0  & 3.9\0  & \no    & \no    \\
\noalign{\hrule}
\noalign{\smallskip}
\end{tabular}
\end{center}
}
\label{tab3}
\tablenotetext{a} {Approximate bin centers in absolute Vega
magnitudes.  The actual first bin center is $-30.15$ for the ``8.0''
columns and $-30.11$ for the ``PAH'' columns.  Successive bin centers
differ by half-magnitude intervals.}
\end{table*}

\clearpage
\begin{table*}[hbt]
{%\scriptsize
\begin{center}
\centerline{\sc Table 4}
\vspace{0.1cm}
\centerline{\sc Schechter Function Parameters}
\vspace{0.3cm}
\begin{tabular}{ccc}
\hline\hline
\noalign{\smallskip}
 & all 8~\micron\ emission & PAH only \\
\hline
\noalign{\smallskip}
$M^*(8~\micron)$ & $-28.46\pm0.08$ & $-28.34\pm{0.08}$ \\
$L^*(8~\micron)$ & $1.8\times10^{10}$ & $1.6\times10^{10}$~\Lsun \\
$\Phi^*$~\tablenotemark{a}         & $1.10\times10^{-3}$  &
               $1.17\times10^{-3}$ \\ 
$\alpha$         & $-1.38\pm0.04$     & $-1.26\pm0.04$ \\
Integrated luminosity\tablenotemark{b} & $3.1\times10^7$ 
& $2.5\times10^7$~\Lsun~Mpc$^{-3}$\\
\noalign{\hrule}
\noalign{\smallskip}
\end{tabular}
\end{center}
}
\label{tab4}
\tablenotetext{a}
{Fiducial galaxy density in galaxies~Mpc$^{-3}$~mag$^{-1}$.
Statistical uncertainty is very small, but the real uncertainty is
about 15\% set by cosmic variance.}
\tablenotetext{b}
{Integral of $\phi(L) dL$, where $L \equiv \nu L_\nu$ for $\nu$
corresponding to 8~\micron.}
\end{table*}

\end{document}